\documentclass[twocolumn,preprintnumbers,amsmath,amssymb,superscriptaddress]{revtex4-1}
\usepackage{graphicx}
\usepackage{dcolumn}
\usepackage{bm}
\usepackage{natbib}

\def\apjs{Astrophys. J. Supp.}
\def\mnras{Mon.Not.Roy.As.Soc.}

\def\ltsima{$\; \buildrel < \over \sim \;$}
\def\simlt{\lower.5ex\hbox{\ltsima}}
\def\gtsima{$\; \buildrel > \over \sim \;$}
\def\simgt{\lower.5ex\hbox{\gtsima}}

\def\[{\begin{equation}}
\def\]{\end{equation}}
\def\m@th{\mathsurround=0pt }
\def\eqalign#1{\null\,\vcenter{\openup1\jot \m@th
 \ialign{\strut\hfil$\displaystyle{##}$&$\displaystyle{{}##}$\hfil
 \crcr#1\crcr}}\,}

\begin{document}
\title{Unmodified Gravity}

\date{\today}
\newcommand{\ud}{\mathrm{d}}
\newcommand{\fpe}{f_\perp}
\newcommand{\fpa}{f_\parallel}
\newcommand{\om}{\Omega_m}

\author{Fergus Simpson}
 \email{frgs@roe.ac.uk}
\author{Brendan Jackson}
 \email{bmj@roe.ac.uk}
\author{John A. Peacock}
\affiliation{SUPA, Institute for Astronomy, University of
Edinburgh, Royal Observatory, Blackford Hill, Edinburgh EH9 3HJ}

\date{\today}

\begin{abstract}
By relaxing the conventional assumption of a purely gravitational interaction between dark energy and dark matter, substantial alterations to the growth of cosmological structure can occur. In this work we focus on the homogeneous transfer of energy from a decaying form of dark energy.  We present simple analytic solutions to the modified growth rates of matter fluctuations in these models, and demonstrate that neglecting physics within the dark sector may induce a significant bias in the inferred growth rate, potentially offering a false signature of modified gravity.

\end{abstract}

\maketitle

\section{Introduction}

One of the major goals of future cosmological studies is to categorise dark energy as one of three candidates: a cosmological constant, a physical fluid, or simply some manifestation of new gravitational physics. Geometric measurements such as supernovae and baryon acoustic oscillations will scrutinise the null hypothesis of a cosmological constant. However, a physical fluid and modified gravity could both reproduce  almost arbitrary expansion histories, so distinguishing them requires a study of the growth of structure. It is the validity of this diagnostic step that provides the focus of the present work.

Anisotropic stress in the dark energy fluid may mimic metric theories of gravity, as demonstrated by Kunz \& Sapone \cite{2007PhRvL..98l1301K} and more generally by Hu \cite{2009NuPhS.194..230H}. Here we take this one step further: even \emph{without} anisotropic stress, it can be shown that significant deviations in structure growth are achievable. The only requirement is that some interaction should exist between dark matter and the smooth dark energy component, permitting the transfer of energy.  

Any change in the dark energy density is conventionally attributed to the equation of state $w(z)$, which dictates the adiabatic behaviour of a physical fluid. But density evolution may also arise from non-gravitational interactions with other fluids, thereby violating adiabaticity. Cosmologies with energy exchange  have been extensively studied in the literature \cite{2000PhRvD..62d3511A, 2006PhRvD..73j3520B, valvi, valvi2, 2009PhRvD..79d3522C, jackson, 2001PhLB..521..133Z, baldi2010, LiBarrow}, and generate an expansion history that is fully reproducible by a single inert scalar field whose evolution matches the effective equation of state $w_{\rm{eff}}(z)$. The degeneracy between these two models may be broken by studying the growth of structure, which is disrupted by the evolving matter density \cite{HeWang}. Yet it is this same test that would conventionally be used to identify modifications to gravity.



In \S \ref{sec:coupling}  we begin by reviewing a simple case of dark energy decaying into a form of dark matter.   The evolution of perturbations are quantified in \S \ref{sec:growth}, before exploring the observational consequences in \S \ref{sec:observables}.

\section{Energy Exchange} \label{sec:coupling}

When speculating on possible interactions amongst the lesser known constituents of our Universe, we are spanning a remarkably broad class of models, with potentially numerous degrees of freedom.   This could extend to new regimes of dark physics, such as  dark matter particles spontaneously decaying into a relativistic dark species such as neutrinos or massless particles, e.g. the dark photons speculated by Ackerman et al \cite{2009PhRvD..79b3519A}. This particular example would be compatible with current observations provided the dark matter particle is sufficiently massive, and the dark fine-structure constant is sufficiently small. For the remainder of this work, we shall focus on the case of dark energy decaying into a form of dark matter, and explore the observational consequences. 

As an illustrative example, we study the simple case of a cosmological (almost) constant that decays into a homogeneous form of dark matter.  The selection $w=-1$ for the dark energy fluid bypasses the various instability issues highlighted in previous work \cite{jackson, valvi, 2009PhRvD..79d3522C}. Instead we choose to focus on the behaviour of the dark matter perturbations.  The evolution of the mean matter density $\rho_m$ is dictated by the conservation equations

\[ 
\rho'_\Lambda = -Q ,
\] 

\[ 
\rho'_m +3 \mathcal{H} \rho_m = Q ,
\]

\noindent where the prime denotes a derivative with respect to conformal time, and $\mathcal{H} \equiv a'/a = \dot{a}$ is the conformal Hubble parameter. The interaction parameter $Q$ controls the rate of energy transfer, and is generally considered to be a function of either the dark matter or dark energy density.
 

The central result of this paper will be to demonstrate that models of interacting  dark energy modify the growth of large scale structure, with an explicit example that exhibits a constant decrement $c$, such that

\[ 
\eqalign{
f & \equiv \frac{\ud\ln\delta}{\ud\ln a} \cr 
&= \Omega_m^{\gamma} - c   , \label{eq:general}
}
\]

\noindent thus providing another mechanism for anomalous growth, aside from modified gravity or anisotropic stress. The magnitude of $c$ is determined by both the nature and strength of the interaction. 

We adopt an empirical modification to the density evolution, which reproduces the functional form for a number of models such as interacting quintessence \cite{2001PhLB..521..133Z, meng}, given by

\[ \label{eq:epsi}
\rho_m = \rho_{m0} a^{-3 + \epsilon}  ,
\]

\noindent where the parameter $\epsilon \ll 1$ dictates the rate of energy transfer.  This particular parameterisation is advantageous in providing the necessary scaling of the dark energy density to help resolve the naturalness and coincidence problems. The dark energy density maintains a magnitude comparable to the dark matter density, although it does not help explain why the transition to dark energy domination has occurred at recent times. Furthermore, the simple form of (\ref{eq:epsi}) allows significant progress to be made analytically. In this case the evolution of the dark energy density is, for a flat universe, given by 

\[ \label{eq:epsi_lambda}
\rho_\Lambda = \rho_{\Lambda 0}  + \frac{\epsilon}{3 - \epsilon} \rho_{m0} \left[a^{-3+ \epsilon} - 1 \right] \, .
\]

At high redshifts, this leads to the tracking behaviour characterised by

\[
\Omega_\Lambda(a \to 0)  = \frac{\epsilon}{3}  ,
\]

\noindent until we approach the era of radiation domination. Thus in effect this model adjusts the dark energy decay lifetime with epoch; for a constant lifetime, we would expect $\rho_\Lambda \to \rm{constant}$ at high $z$ and  $\rho_\Lambda \to 0$ at late times. Our analysis treats the parameter $\epsilon$ as a constant, though remains valid provided $\epsilon$ varies sufficiently slowly, satisfying the condition $\ud \epsilon / \ud \ln a \ll \epsilon$.

Given the dark matter mass evolution (\ref{eq:epsi}), one can see a direct relation between our phenomenological approach, and that of a coupled scalar field as outlined in \cite{2000PhRvD..62d3511A}.

\[
\epsilon_{\rm{eff}} = \frac{-\int Q (\phi) \ud \phi} {\ln a}  .
\]

\section{Growth of Structure} \label{sec:growth}

Any form of non-gravitational interaction in the dark sector may be expected to impact upon the growth of structure. As we shall see, a decaying form of dark energy generates three distinct mechanisms for slowing the rate of structure growth. We quantify each of these in turn, starting with the modification to the expansion history $H(z)$.


\subsection{Background Dynamics}

Here we generalise the derivation of linear structure growth by Linder \& Cahn \cite{2007APh....28..481L} to incorporate the different background evolution of both dark energy and dark matter, $\rho_m \propto
a^{-3+\epsilon}$.  Including energy exchange at the
background level \emph{only} is insufficient, as we will show, but it is instructive to identify the contribution this yields in the following derivation of $f(a) \equiv \ud \ln \delta / \ud \ln a$. Starting from the differential equation

\[ \label{eq:diffy}
\frac{\ud^2 \delta}{\ud t^2} + 2H(a) \frac{\ud \delta}{\ud t} - 4 \pi \rho_m \delta = 0 ,
\]

\noindent which may be rewritten as

\[
\eqalign{
\frac{\ud f}{\ud \ln{a}} &+ 
 \frac{1}{2} \frac{\ud \ln{H^2}}{\ud \ln{a}} (f-\frac{1}{2}) + f(f+2)
 \cr & -
\frac{3}{2}\om(a) = 0 \, ,}
\]
and utilising

\[
H^2/H_0^2 \simeq \om a^{-3+\epsilon}\left[1+ \Omega_\Lambda(a)/\om(a) \right] \, ,
\]
yields, to first order in $\left(f(a) - 1\right)$ \cite{2007APh....28..481L},

\[ \label{eq:G}
\eqalign{
f(a) \approx 1 - \frac{\epsilon}{5}   &-
\frac{1}{2}\Omega_\Lambda(a)\left(1-\frac{\epsilon}{5}\right) \cr &+
\left[1-\frac{1}{2} \Omega_\Lambda(a) \right]I(a) 
\, ,
}
\]
where
\[
I(a) = -\frac{1}{4} 
 a^{-(5+\epsilon)/2} \int^a_0 \frac{\ud a'}{a'} (a')^{(5 + \epsilon)/2} \Omega_\Lambda(a') \,.
\]
Before evaluating this integral, we first need to consider the
scaling behaviour of dark matter and dark energy at early times. To proceed, we simply recast the dark energy density (\ref{eq:epsi_lambda}) in the form
\[
\Omega_\Lambda(a) = \, \Omega_1 + \Omega_2(a) + O\left(\epsilon^2\right) +  O \left(\Omega_\Lambda^2(a) \right) \, ,
\]
where $\Omega_1$ is the constant component, and $\Omega_2$ is the collection of terms that follow a power law:
\[
\eqalign{
\Omega_1 &= \frac{\epsilon}{3}; \cr 
\Omega_2 &= \left(\Omega_{\Lambda 0} - \frac{\epsilon}{3}\right) a^{3-\epsilon} \, , 
}
\]

\noindent provided we assume the contribution from radiation is negligible.

Using these redshift dependencies in $I(a)$, then to first order in deviations from matter domination we find:
\[
I(a) = -\frac{1}{4} \left[\frac{2}{5+\epsilon} \Omega_1 +  \frac{2}{11 - 3\epsilon} \Omega_2(a) \right] \,.
\]
Writing $\Omega_2(a) = \Omega_\Lambda(a) - \Omega_1$ leaves us with
\[ \label{eq:I}
I(a) =  - \frac{\epsilon}{30} + \frac{\epsilon}{66} - \frac{\Omega_\Lambda(a)}{22}  + O\left(\epsilon^2\right) +  O \left(\Omega_\Lambda^2(a) \right) \, .
\]
Substituting our solution back into (\ref{eq:G}) gives:
\[
f(a) \simeq 1 - \left(\frac{6}{11} - \frac{6}{55} \epsilon \right) \Omega_\Lambda(a) - \frac{12}{55} \epsilon
\]
Of these two new terms involving $\epsilon$, the second is of greater importance. This contrasts with the conventional approximation for $f(a)$ \cite{1998ApJ...508..483W, 2005PhRvD..72d3529L} given by

\[
\eqalign{
f(a) &= \Omega_m^{\gamma}(a) \cr
& \simeq  1 - \gamma \Omega_\Lambda (a) \, .
}
\]

\noindent If we insist on maintaining the definition $\gamma \approx \left[1  - f(a) \right]/\Omega_\Lambda(a)$, the extra $\frac{12}{55} \epsilon$ will mean $\gamma$ is no longer a constant to first order in $\Omega_\Lambda(a)$. This suggests that a constant $\gamma$ is no longer a viable approximation for solving the perturbation equations.
Rather than face the difficulty in generalising to a redshift
dependent $\gamma$, we note that the constancy of $\gamma$ can be
maintained at the appropriate level of approximation by simply
changing the approximate solution such that
\[ \label{eq:growth2}
g(a) = e^{{\int^a_0 \left(\ud a'/a'\right)\left[\om(a')^\gamma - 1 + b\right]}} \,.
\]
The new constant $b$ means that
\[
\gamma \approx - (f(a) - 1 + b)/\Omega_\Lambda(a) \,,
\]
and we see that if $b$ is chosen to offset the
constant terms in $(f(a)-1)$, then $\gamma$ will once again be a constant. 
For the case of pure background energy exchange,

\[
b = - \frac{12 \epsilon}{55} \, ,
\]

\noindent and hence our first modification to the growth rate is

\[
f(a) \simeq \Omega_m^{\gamma}(a) - \frac{12 \epsilon}{55} \,  ,
\]

\noindent where $\gamma$ is subject to a minor perturbation

\[ \label{eq:gammapert1}
\gamma = \frac{6}{11} - \frac{6}{55} \epsilon \, .
\]

\subsection{Dilution}

The energy exchange does not only affect the background dynamics; it
also generates new terms in the perturbation equations, so (\ref{eq:diffy}) is no longer valid.  By adding a homogeneous contribution to the background density, the fractional contrast, $\delta$, is suppressed. In this particular parameterisation of energy exchange, the density and velocity perturbation equations are greatly simplified from the general case presented in \cite{valvi}:

\[ \label{eq:delta0}
\delta' + \epsilon \mathcal{H} \delta + \theta - 3\Phi' = 0  ;
\]

\[ \label{eq:theta0}
\theta' +  \mathcal{H} \theta - k^2 \Psi = 0  .
\]

\noindent The differential equation governing $f(a)$ now becomes

\[
\eqalign{
\frac{\ud f}{\ud \ln{a}} &+ 
\frac{f}{2} \frac{\ud \ln{H^2}}{\ud \ln{a}} + f(f+2) + 2\epsilon
 \cr & +  \frac{\epsilon}{2} \frac{\ud
  \ln{H^2}}{\ud \ln{a}} - \frac{3}{2}\om(a) = 0}
\]
and carrying these new terms through gives
\[
\eqalign{
f(a) \approx 1-\frac{2}{5} \epsilon  + \left[1-\frac{1}{2}
\Omega_\Lambda(a) \right](1-4\epsilon)I(a)  \,.
}
\]
The integral $I(a)$ has already been evaluated (\ref{eq:I}), and so we find:
\[
f(a) \approx 1 + \left(\frac{6}{11} - \frac{16}{55} \epsilon \right) \Omega_\Lambda(a) - \epsilon \left(\frac{2}{5} + \frac{1}{55} \right) \,,
\]
or 
\[
f(a) = \Omega_m^{\gamma}(a) - \frac{23}{55} \epsilon \, ,  
\]
where
\[
\gamma = \frac{6}{11} - \frac{16}{55} \epsilon  \, .
\]

\begin{figure}[t]
\includegraphics[width=80mm]{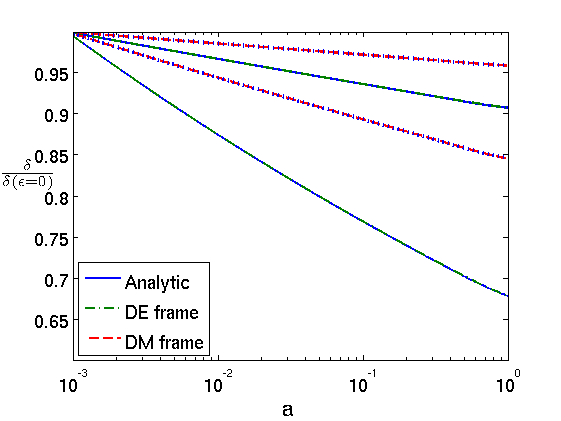}
\caption{The fractional change in the evolution of linear perturbations in the presence of a decaying cosmological constant with $\epsilon=0.01$ (upper) and  $\epsilon=0.04$ (lower), as defined in (\ref{eq:epsi}).  The pair of dash-dotted lines highlight the slight suppression of growth induced by the diluting effect of transferring homogeneous energy into the dark matter frame. The dashed lines correspond to an energy transfer in the CMB rest frame, with the extra deceleration arising due to the introduction of stationary matter. These are well described by the solid lines, which illustrate the analytic solution given by (\ref{eq:ana}).}
\label{fig:fig1}
\end{figure}

\begin{figure}[t]
\includegraphics[width=80mm]{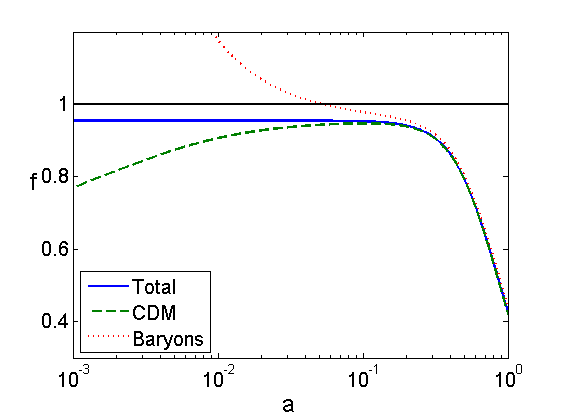}
\caption{The logarithmic growth rate of both cold dark matter (dashed), and the baryons (dotted), in the presence of a decaying cosmological constant with $\epsilon=0.04$. The solid line illustrates the approximate solution given by (\ref{eq:ana2}), which traces the total matter perturbation.}
\label{fig:fig1b}
\end{figure}

\subsection{Inertial Drag}

In the previous subsection we implicitly assumed that there is a pure energy transfer, from the perspective of the dark matter rest frame. If however the energy transfer occurs in the CMB frame, the newly formed dark matter will not be instilled with the appropriate bulk motion, and a drag term appears in the perturbation equations. 

The density and velocity perturbation equations are then of the form
\[ 
\label{eq:delta}
\delta' + \epsilon \mathcal{H} \delta + \theta - 3\Phi' = 0  ,
\]

\[ \label{eq:theta}
\theta' +  \mathcal{H} \left( 1 + \epsilon \right) \theta - k^2 \Psi = 0  .
\]

The physical interpretation of the $\epsilon  \mathcal{H} \delta$ term in (\ref{eq:delta}) is one of dilution, as addressed in the previous subsection. A second effect which also slows structure growth is the extra drag term in (\ref{eq:theta}), $\epsilon  \mathcal{H} \theta$. This arises due to the appearance of stationary matter, reducing the mean flow rate.

If we define new perturbations $\bar{\delta} = \delta a^{\epsilon}$
and  $\bar{\theta} = \theta a^{\epsilon}$, 
then substituting these rescaled variables into (\ref{eq:delta}) and (\ref{eq:theta}) leaves us with
\[ 
\label{eq:delta_bar}
\bar{\delta}' + \bar{\theta} = 0 \, ,
\]
\[  \label{eq:theta_bar}
\bar{\theta}' +  \mathcal{H} \bar{\theta} - k^2 \bar{\Psi} = 0  \, .
\]
\noindent where we have neglected $\Phi'$, and the gravitational source term is rescaled as
\[ -k^2 \bar{\Psi} = \frac{3}{2} \mathcal{H}^2 \Omega_m \bar{\delta}  
= \frac{3}{2} \mathcal{H}^2 \Omega_m \delta a^{\epsilon} =  -k^2
\Psi a^{\epsilon}   \,.
\]
These two new equations, (\ref{eq:delta_bar}) and
(\ref{eq:theta_bar}), are precisely the equations for perturbations as
if no energy exchange were taking place beyond a background level. The previous section provided us with an approximation solution in a
universe with background energy exchange:
\[
\bar{\delta}' \simeq \bar{\delta} \mathcal{H}
\left[\Omega_m(a)^{\gamma} - \epsilon \left(\frac{1}{5} +
  \frac{1}{55} \right) \right]  \, .
\]
After rescaling:
\[ 
\delta' \simeq \delta \mathcal{H} \left[\Omega_m(a)^{\gamma} -
  \epsilon \left(\frac{6}{5} + \frac{1}{55}\right) \right] \, ,
\]
or simply
\[ 
\label{eq:ana} 
f(a) = \Omega_m^{\gamma}(a) - \frac{67}{55} \epsilon \, ,
\]
\noindent and we return to the solution (\ref{eq:gammapert1}) to find
\[ 
\gamma = \frac{6}{11} - \frac{6}{55} \epsilon \, .
\]

The physical mechanisms contributing to this new growth rate are summarised in Table \ref{table:table1}. The quoted total corresponds to the model where Q is unperturbed, and with the energy transfer occurring in the dark energy rest frame (that of the CMB). If instead the dark energy decay is sensitive to local fluctuations in the matter density, $\delta Q \propto \mathcal{H} \delta \rho_m$, the dilution term is no longer present. Similarly, if the energy exchange were to take place in the dark matter rest frame as opposed to the dark energy rest frame, the drag contribution vanishes.

\begin{table}
\begin{tabular*}{50mm} {@{\extracolsep{\fill}} l  c  c }
\hline
Contribution & $\Delta f / \epsilon $ & $\Delta \gamma / \epsilon$ \\ 
\hline \\  
Background & $\displaystyle - \left( \frac{1}{5} + \frac{1}{55} \right)$  & $- \displaystyle \frac{6}{55}$ \\ 
\noalign{\smallskip} \hline \\ 
Dilution & $- \displaystyle\frac{1}{5}$ &  $-\displaystyle\frac{10}{55}$  \\ 
\noalign{\smallskip} \hline \\ 
Drag & $- \displaystyle\frac{4}{5}$ &  $\displaystyle + \frac{10}{55}$\\
\noalign{\smallskip} \hline \\ 
Total & $\displaystyle - \frac{67}{55}$ & $- \displaystyle \frac{6}{55}$ \\ 
\noalign{\smallskip} \hline  
\end{tabular*}  
\caption{Summary of the contributions to the growth rate from a decaying dark energy component, in units of $\epsilon$. Whether the effects of dilution and drag arise will depend on the behaviour and orientation of the interaction parameter $Q$. Note that the terms in the left column, those contributing directly to $f(z)$, have a considerably greater impact than the corrections to $\gamma$ in the right column.} \label{table:table1}
\end{table}

\begin{figure}[t]
\includegraphics[width=80mm]{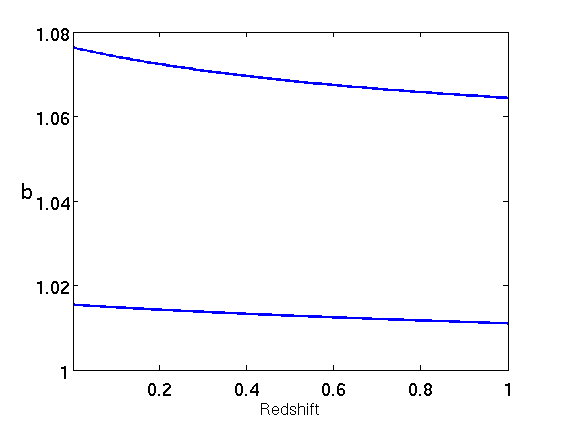}
\caption{The baryonic bias induced by $\epsilon=0.01$ and $\epsilon=0.04$.}
\label{fig:bias}
\end{figure}

\subsection{Baryonic Correction}

Thus far we have assumed the universe comprises solely of dark matter and dark energy, but before considering the observational consequences we must first extend our analysis to include baryons. We therefore divide the matter component into cold dark matter and baryonic terms, denoted $\rho_c$ and  $\rho_b$. The baryonic perturbations are simply given by
\[
\delta''_b +\mathcal{H} \delta'_b = 4 \pi G a ( \rho_c \delta_c +\rho_b \delta_b) .
\]

 The analytic solution in (\ref{eq:ana}) now requires a minor rescaling to compensate for the introduction of baryonic matter. A simple weighting of the decay parameter $\epsilon$ by the fraction of mass to which it applies is sufficient.
\[ \label{eq:ana2}
f = \Omega^{\gamma}_m - \frac{67}{55} \bar{\epsilon}  ,
\]

\[
\bar{\epsilon} \equiv \frac{\Omega_{c}}{\Omega_{m}} \epsilon  .
\]

In Figure \ref{fig:fig1b}, this formalism is seen to provide an excellent description for the combined matter perturbation. Although we note that a more thorough treatment of the perturbations is required at very high redshifts ($z>100$), where the radiation energy density starts to become significant. At later times the baryons closely track the growth rate of the dark matter perturbations, but with a slightly enhanced density contrast. Defining the baryonic bias $b \equiv \delta_b/\delta_c$, the present day value of this ratio is approximately given by 
\[
b \approx  1 + 2 \epsilon  ,
\]
\noindent as illustrated in Figure \ref{fig:bias}.

\begin{figure}[t]
\includegraphics[width=80mm]{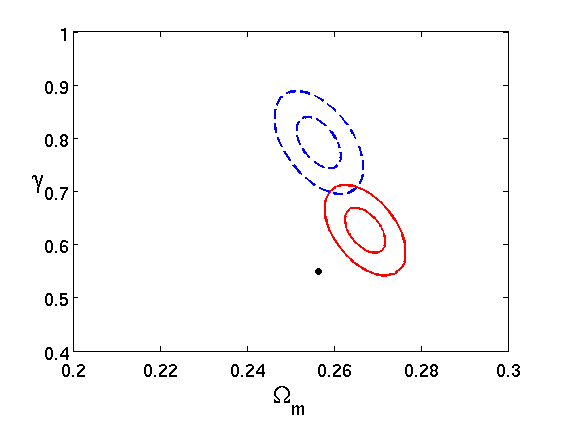}
\caption{The solid contours provide an example of the bias which may be induced in the gravitational growth index $\gamma$ when making the false assumption that dark energy is stable.  The true model, as indicated by the black dot, corresponds to $\epsilon=0.01$. The dashed contours demonstrate the modification to the growth index induced by the elastic interaction model outlined in \cite{simpscat}.}
\label{fig:fig4}
\end{figure}

\section{Observational Consequences} \label{sec:observables}

We now review their impact on cosmological observables, using redshift-space distortions as a measure of the growth rate, before considering implications for the Integrated Sachs Wolfe (ISW) effect.

\subsection{Redshift Space Distortions}

In order to highlight the observational consequences of this energy exchange, we evaluate the appropriate Fisher matrix for a redshift survey at $z=0.5$ (see \cite{simp09}), and combine this with the DETF Fisher matrix for Planck. We marginalise over the parameter set \[
[w_0, w_a, \Omega_\Lambda,  \Omega_k, \Omega_m h^2, \Omega_b h^2, n_s, A_s, \beta, \gamma,
  \sigma_p, \epsilon].
\]
The standard cosmological parameters are taken to have fiducial values as
derived from WMAP5 \cite{2008arXiv0803.0547K}. In order to remain consistent with the CMB fisher matrix, when perturbing $\epsilon$ we ensure the value of $\rho_m$ at $z=1100$ is held fixed. The Hubble parameter $h$ is slightly perturbed for the purposes of distance estimation. We neglect the small change induced by the Integrated Sachs Wolfe effect (see \ref{sec:ISW}).

As a parameterisation of modified gravity, the growth of structure is taken to follow the form given by \cite{1998ApJ...508..483W, 2005PhRvD..72d3529L} 

\[ \label{eq:gindex}
f \simeq \Omega_m^{\gamma} .
\]

Clearly one ought to expect that, for the case of interacting models, applying the above prescription would lead to a biased estimate of the growth index $\gamma$. In addition to the extra $\epsilon $ term in (\ref{eq:ana}), the value of $\gamma$ is biased if the evolution of $\Omega_m(z)$ does not run as expected. This shift is illustrated by the solid contours in Figure \ref{fig:fig4} which are generated with $\epsilon=0.04$. 

In practice, we must deal with the Fingers of God in greater detail than a single parameter $\sigma_p$. It should also be noted that these models may have some impact on virialised structures, though we leave this as a topic for future investigation.

\subsection{Integrated Sachs Wolfe Effect \label{sec:ISW}}

One restriction that dark energy models must satisfy is not to overpredict the Integrated Sachs Wolfe effect. Excessive change in the gravitational potential would invariably generate large fluctuations in both the CMB and CMB-LSS cross-correlation on large angular scales. Indeed this has already been employed by Valiviita et al. \cite{valvi2} to assist in constraining models of decaying dark energy.

Since we are still working in the context of General Relativity, the gravitational potential is readily evaluated via the Poisson equation

\[
\Phi(k, a) = - 4 \pi G \rho(t) a^2 \frac{\delta(a)}{k^2} \, ,
\]

\[
\Psi = \Phi \, .
\]

\begin{figure}[t]
\includegraphics[width=80mm]{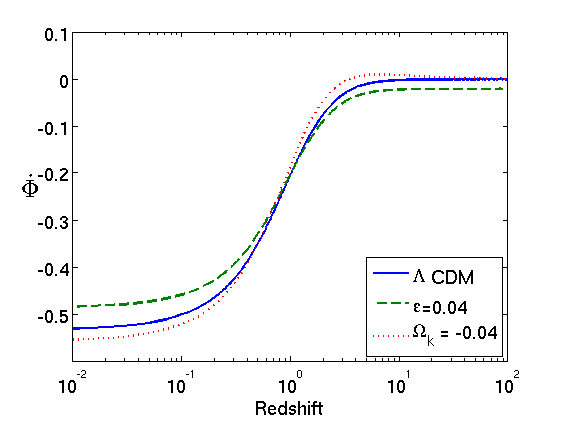}
\caption{The decay rate of the gravitational potential as experienced in a flat $\Lambda \rm{CDM}$ Universe (solid), and with perturbed cosmologies $\epsilon=0.01$ (dashed) and $\Omega_k=-0.01$ (dotted).}
\label{fig:isw}
\end{figure}

\noindent This evolution is illustrated in Figure \ref{fig:isw} for the standard case of flat $\Lambda \rm{CDM}$, alongside small perturbations in the decay parameter $(\epsilon)$  and global curvature $(\Omega_k)$. While there is a distinctive change in behaviour at high redshift, this is a regime that lies out of reach for cross-correlation studies. Radiation may also start to become significant at this point, we have not included this effect in our treatment.

The ISW signal is, at best, a $\sim 5 \sigma$ observation, and as such there is significant room for flexibility in the anticipated signal strength. Taking a $20\%$ change as the maximum permissible, this corresponds to an approximate upper limit of $\epsilon \lesssim 0.1$. This appears broadly consistent with the findings of Valiviita et al. \cite{valvi2}, who established an upper bound of $|\Gamma| < 0.23 H_0$ for a constant dark energy decay rate, from a combination of WMAP, supernovae, and BAO data. The two parameterisations are related by $\Gamma = - \mathcal{H} \epsilon$. Our earlier analysis on the growth rate may be naturally extended to this model, which we also find to be well described by the prescription

\[
\Delta f \propto \Gamma/\mathcal{H} \, .
\]

\section{Discussion}

In the case of a decaying cosmological `constant', we have demonstrated that the growth rate of large scale structure is subject to a constant decrement $f=\Omega_m^{\gamma} - c$. For the simple model we consider, this can largely be attributed to the dragging effect on bulk motions induced by the production of stationary matter. Smaller contributions arise from both the background dynamics, and the diluting effect of gradually introducing a homogeneous density field. It is also important to note that in more general interacting models the presence of dark energy perturbations may also influence the growth of structure. These act to enhance the growth rate, and may overpower the mechanisms considered here. Indeed in many cases the dark energy perturbations in coupled models lead to pathological instabilities.

Energy exchange within the dark sector leads to a change in the comoving matter density, which in turn invokes a number of changes to cosmological observations. If these changes are not taken into consideration, then suppression to the growth of large scale structure leads to a na\"{i}vely inferred value of the growth index $\gamma$ rising above the conventional value $\gamma > 0.55$. While the physical motivation for such models remains unclear, we believe this is no less true for current approaches to modified gravity.

As noted by Blandford et al. \cite{2005ASPC..339...27B}, a blind cosmologist living in the radiation dominated era might measure the evolution of the scale factor and erroneously conclude that the expansion is driven by a scalar field with an exponential potential. This thought experiment may be extended further: a ``dark" astronomer only capable of studying the dark matter would notice a strange behaviour of non-linear structure. They might attempt to construct a modified theory of gravity which accounts for this behaviour. However the true source of this discrepancy is the momentum exchanged between the baryons and photons, inducing new features to the growth of structure.  This modification is apparent both at the era of recombination, and the present day.
We are at risk of falling into a similar trap here: the baryocentric assumption that neither dark energy nor dark matter exhibit any complex behaviour may lead to dark physics being mistaken for modified gravity.

\noindent{\bf Acknowledgements} \\
FRGS and BMJ are grateful for support from the STFC, and we thank the anonymous referee for helpful comments.

\bibliography{/Volumes/katrine.roe.ac.uk/Routines/dis}


\end{document}